\documentclass[twoside,a4paper,fleqn]{svjour2}                    % onecolumn
\smartqed  % flush right qed marks, e.g. at end of proof
\usepackage{graphicx}\usepackage{amsmath,mathrsfs}
\usepackage{epsfig}\usepackage{psfrag}
\usepackage{subfigure}
\usepackage{tabularx}\usepackage{rotating}
\def\figurename{Fig.}

\addtocounter{footnote}{0}

 \usepackage{mathptmx}      % use Times fonts if available on your TeX system
\usepackage{latexsym}
 \journalname{Celestial Mechanics and Dynamical Astronomy}
\begin{document}
%\pagestyle{myheadings}
%\markboth{Z.~Jiang \& L.~Ossipkov}{Anisotropic distribution functions for spherical galaxies}

\title{Dynamical modelling of the elliptical galaxy NGC~2974%\thanks{Grants or other notes
}

\titlerunning{Dynamical modelling of the elliptical galaxy NGC~2974}        % if too long for running head

\author{Zhenglu Jiang}

\authorrunning{Z.~Jiang}

\institute{Z.~Jiang \at
              Department of Mathematics, Zhongshan University,
              Guangzhou 510275, China \\
              \email{mcsjzl@mail.sysu.edu.cn}
}

\date{Received: \today / Accepted: date}

\maketitle

\begin{abstract}
 In this paper we analyse the relations between a previously described
oblate Jaffe model for an ellipsoidal galaxy
and the observed quantities for NGC~2974, and obtain
 the length and velocity scales for a relevant elliptical galaxy model.
We then derive the finite total mass of the model from
these scales, and finally find a good fit of an isotropic oblate
Jaffe model
by using the Gauss-Hermite fit
parameters and the observed ellipticity of the galaxy NGC~2974.
The model is also used to predict the total luminous mass of NGC~2974,
assuming that the influence of dark matter in this galaxy on the
image, ellipticity and Gauss-Hermite fit
parameters of this galaxy
is negligible within the central region, of radius $0.5R_{\rm e}.$

\keywords{celestial mechanics \and stellar dynamics \and galaxies}
\end{abstract}

\section{Introduction}
\label{intro}
Galaxies are usually modelled in terms of their morphology,
surface luminosity profile, mass-light ratio,
velocity dispersion profile and the line-of-sight velocity distribution curve
(often termed the rotation curve). Recently, dynamical modelling
of the ellipsoidal galaxy NGC~2974 was performed by
  Cinzano and van der Marel (1994), based on the Gauss-Hermite
fit parameters, i.e. the rotation
velocities $\hat{v},$ velocity dispersions $\hat{\sigma}$
and deviations of the velocity
profiles from a Gaussian, as quantified by the Gauss-Hermite moments
$h_3$ and $h_4.$ A similar modelling of the ellipsoidal galaxy M32 was
 given by Qian et {al.} (1995). After that, Baes
and Dejonghe (2004) provided a
 completely analytical class of dynamical models for spherical
galaxies and bulges with central black holes.  Baes et
al.~(2005) also studied the dynamical structure of isotropic spherical
galaxies with a central black hole.
Some dynamical models of luminous and dark matter for 17
early-type galaxies were recently presented by Thomas et {al.}
(2007). Previously a number of more extensive studies have been published
(e.g. Gerhard 1993a,b; van der Marel \& Franx
1993; Bender et {al.} 1994; Gerhard et {al.} 1998; Kronawitter et
{al.} 1999; Saglia et {al.} 2000; Halliday et {al.} 2001). In
particular, the works of Emsellem, Goudfrooij and Ferruit (2003) and
Krajnovi\'{c} et al.~(2005) are very thorough investigations of NGC~2974.
The former shows a two-arm gaseous
spiral in the inner 200 pc of the early-type galaxy NGC~2974 by
fitting the TIGER and long-slit stellar kinematic data of the galaxy
with a dynamical model, not requiring the addition of a disc or a
central dark mass.  For their dynamical model,
a numerical two-integral distribution function
was derived from a multi-Gaussian expansion
mass model by using the Hunter-Qian (1993) algorithm. The paper of
Krajnovi\'{c} {et al.} presents
a very thorough analysis of NGC~2974 using completely different data
and methodology. Krajnovi\'{c} {et al.} first obtained a
multi-Gaussian expansion mass density, fitting simultaneously the
ground-based $I\hbox{-band}$ image and dust-corrected PC part of the
WFPC2/F814 photometry of this galaxy; they then constructed a
three-integral model from the recovery of the mass density using the
Schwarzschild (1979) orbit-superposition method; finally, they
provided a dynamical modelling of stars and gas in this galaxy by use
of the Gauss-Hermite fit parameters.
 In this paper we again model dynamically NGC~2974, by
employing distribution functions from self-consistent
  density-potential pairs of
an oblate Jaffe model (Jiang 2000) and
using the Gauss-Hermite fit
parameters and the ellipticity (Sandage and Tammann 1981).
The line-of-sight velocity distribution curves are
calculated from the distribution functions for the oblate Jaffe model
and quantified by its Gauss-Hermite fit parameters.
The derived Gauss-Hermite fit parameters for the oblate Jaffe model
correspond very well to
 those of NGC~2974. The model is also applied to predict
the total luminous mass of NGC~2974, since in the region of a radius
of $0.5R_{\rm e}$ ($R_{\rm e}=33.7$ arcsec) the dynamical influence
of the dark halo of NGC~2974 is negligible (cf.~Cinzano and van der
Marel 1994).

As indicated above, a distribution function can be derived from a mass
density by either the Hunter-Qian algorithm or the Schwarzschild
method. As in some other published studies (e.g. Qian et {al.}~1995;
Emsellem, Goudfrooij and Ferruit 2003), we obtain below a
distribution function using the Hunter-Qian (1993) algorithm. It is
worth mentioning that the first step of our dynamical model is to
give directly the oblate Jaffe mass model by using only the galactic
ellipticity in order to  mimic the image of NGC~2974.
This is different from the approach used by
all the earlier works mentioned above.

NGC~2974 has been studied frequently
(e.g. Gallouet et al.~1975;
Sandage and Tammann 1981; Demoulin-Ulrich et al.~1984; Forman et al.~1985;
Canizares et al.~1987; Davies et al.~1987; de Vauouleurs et al.~1991).
It appears to be completely isolated (Kim et al.~1988), is
classified as E4 by Sandage and Tammann (1981), and has a
regular distribution of gas (Kim et al.~1988) and
a dark halo (Cinzano and van der Marel 1994).
The total mass of NGC~2974 within a galacto-centric radius of $13$  kpc
is estimated as $3.8\times 10^{11}$ $h^{-1}$M$_\odot$
by Kim et al.~(1988), and its total luminous mass as
$1.2\times 10^{11}$ $h^{-1}$M$_\odot$
(Cinzano and van der Marel 1994), where $h$
is Hubble's constant. A range of a factor of 2 is given for
$h$ [$h$ is in units of
$100$ kms$^{-1}$Mpc$^{-1}$ and is thought to lie
in the range  $0.5\sim 1$].
Since the first results of the WMAP mission have been made public,
we now know the constant $h$ with a much smaller uncertainty (Bennett 2003).
 Cinzano and van der Marel (1994) also
showed that the total mass of the dark halo is about
$1.8\times 10^{11}$ $h^{-1}$M$_\odot.$
Thus, according to this estimate, the total mass of NGC~2974
 is $3.0\times 10^{11}$ $h^{-1}$M$_\odot,$
which is obviously smaller than that given by Kim et al., mentioned above.
 The difference appears to be due mainly to differences in the
adopted distance and in minor details of the model
(van der Marel, private communication).
Kim (1988) showed that
the optically measured heliocentric velocity  of NGC~2974 is
$1924$ kms$^{-1}$
according to the optical definition $v=c\Delta\lambda/\lambda_0,$ and that
the distance to NGC~2974 is given in terms of the value of Hubble's constant
and is $22.5$ $h^{-1}$Mpc in terms of
the Virgocentric flow model of
Aaronson et al.~(1982) [it is $19.25$ $h^{-1}$Mpc
based on the  Hubble law]. The rotational velocity of NGC~2974
 is about $360$  kms$^{-1}$
(Kim et al.~1988)
and its central velocity dispersion is about $222$  kms$^{-1}$
(Kim et al.~1988; Cinzano and van der Marel 1994).
Gauss-Hermite fit parameters of NGC~2974
along the major and minor axes have been
derived  independently by Bender et al.~(1994) and Cinzano and van der Marel (1994).

We use a stellar dynamical galaxy model
to calculate Gauss-Hermite fit parameters
$(\hat{\sigma},\hat{v},h_3,h_4)$ as a function of the dimensionless
position variable $x^\prime$ (measured from the potential centre
along the major axis).
 The physical parameters in the models
can then be related to the physical dimensions of the observed galaxy
with the help of the length and velocity scales  determined
by minimizing  a certain function. This function can be
constructed  in the physical domain by use of dimensionless
Gauss-Hermite fit parameters
$(\hat{\sigma},\hat{v},h_3,h_4),$ together with the corresponding observations
of an elliptical galaxy,
by using a weighted least squares method.
An isotropic model can then be selected from this class of
axisymmetric elliptical galaxy models by analysing
the image and ellipticity of an elliptical galaxy (in our case NGC~2974).
It is known from Fig.~\ref{NGC2974_MJ_60} in Sect.~\ref{fojm} that the
 Gauss-Hermite fit parameters
$(\hat{\sigma},\hat{v},h_3,h_4)$ of the model can fit
the corresponding  observations of the elliptical galaxy NGC~2974
in terms of the above least squares method.
Finally, the total mass of the best fit model of NGC~2974 can be derived from
the length and velocity scales obtained by comparing
the Gauss-Hermite fit parameters
$(\hat{\sigma},\hat{v},h_3,h_4)$ with the corresponding observations
of NGC~2974.

The models we give here are ideal,
in that effects from the surrounding environment are not considered.
The more a galaxy is  affected by neighbouring objects,
the less well the observations can be expected to be fitted by
such an ideal model. A galaxy needs to be quite close to us for
 adequate observations to be available to determine such effects.
It is not obvious that the best fitting model will be a flattened Jaffe model,
but here we restrict ourselves to the latter.
We first construct the oblate Jaffe model and
search for a fit. Obviously, the existence of a good fit is not
guaranteed but it turns out that one can be found.
It is possible that other such `suitable' galaxies amenable to this
analysis exist but,
up to now, no suitable galaxy other than NGC~2974 has been found.

This paper is arranged as follows.
We first introduce oblate Jaffe models and Gauss-Hermite fit parameters
in Sect.~\ref{ojmgfp}. Then, in Sect.~\ref{rmov},
we mainly analyse relations between the oblate Jaffe model
and observed quantities,
in order to obtain the length and velocity scales of the oblate models.
Having derived the finite total mass of the models from
these scales in Sect.~\ref{dtmm}, we can then, based on the Gauss-Hermite fit
parameters and the ellipticity of
  an elliptical galaxy NGC~2974,
find an isotropic oblate
Jaffe model in Sect.~\ref{fojm} that is a good fit.
The derived total mass of the isotropic model
only reflects the total luminous mass of NGC~2974. This
is because we can neglect the influence of dark matter in NGC~2974 on the
image, ellipticity and Gauss-Hermite fit
parameters of the observed galaxy within the radius of $0.5$ $R_{\rm e}.$
Our work is summarized in Sect.~\ref{concl}.

\section{Oblate Jaffe Models and Gauss-Hermite Fit Parameters}
\label{ojmgfp}
Jiang (2000) defined a class of
oblate axisymmetric elliptical systems with the potential-density pairs
\begin{equation}
\Phi(R^2,z)=\frac{GM}{r_J}
\ln\left(\frac{\sqrt{R^2+(\sqrt{z^2+c^2}+d)^2}}
{\sqrt{R^2+(\sqrt{z^2+c^2}+d)^2}+r_J}\right)\label{(1.13)}
\end{equation}
and
\begin{equation}
\rho(R^2,z)=\frac{M}{4\pi r_J}\frac{c^2dr_J\tau^3+r_J^2\tau^2(Y^3+c^2d)
+(3\tau+2r_J)c^2r_JY(Y+d)^2}{\tau^4(\tau+r_J)^2Y^3},
 \label{(1.14)}
\end{equation}
where $R$ and $z$ are two of three
cylindrical coordinates $(R,z,\phi),$
$M,$ $c,$ $d$ and $r_J$ are positive constants,
 $G$ is the gravitational constant,
$Y=\sqrt{z^2+c^2}$ and $\tau=\sqrt{R^2+(Y+d)^2}.$  Here, these systems are obtained
by flattening the spherical Jaffe (1983) model and they are oblate, and so they
are called oblate Jaffe models.
Some more general extensions of the spherical Jaffe model
have been given by Jiang and other coauthors (Jiang and Moss 2002;
 Jiang, Fang and Moss 2002).
It is worth mentioning that the spherical Jaffe model is a special case of a family of spherical
$\gamma$ models (Kuzmin, Veltmann, Tenjes 1986;
Dehnen 1993; Saha 1993; Tremaine et al.~1994) which have some similar flattening properties
(Jiang, Fang, Liu and Moss 2002; Jiang and Ossipkov 2006; Ossipkov and Jiang 2007a,b).
In the literature, there are also other spherical models
(e.g. Veltmann 1961, 1965, 1979, 1981; Kuzmin and Veltmann 1967a,b, 1973;
Ossipkov 1979a,b; Binney and Tremaine 1987)
and generalized systems with three-dimensional gravitational potentials (e.g. Kutuzov and Ossipkov 1981).

 It is known that
the density (\ref{(1.14)}) at large distances
decays radially like $r^{-4},$ except on the major axis, and as
 $r^{-3}$ on the major axis, where $r$ is one of three
spherical coordinates $(r,\theta,\phi).$
 The central ratio $\alpha$
of isodensity contour $R$-axis to $z$-axis (Jiang 2000)
is given by
\begin{equation}
\alpha^2=\frac{(c+d)(15c^3+27c^2d+15cd^2+3d^3+10c^2r_J+9cdr_J+3d^2r_J)}
{c^2(5c+d)(3c+3d+2r_J)}. \label{(2.5)}
\end{equation}
The total mass of the model is finite and equal to $M.$

Jiang (2000) also proved that the form of the  potentials allows
their two-integral even distribution functions to be calculated
from the self-consistent density-potential pairs
of the above oblate Jaffe model by
using a modification of Hunter and Qian's algorithm (Hunter and Qian 1993;
Qian et al.~1995).
 Once the two-integral even distribution functions have been obtained,
the distribution functions $f(\varepsilon,L_z)$ can be found
by using the maximum entropy principle (Dejonghe 1986), where
$\varepsilon$ and $L_z$ are the relative energy and the $z$-axis
angular momentum respectively.
 For oblate stellar models,
other distribution functions can also be found depending on whether
  isotropy or anisotropy (Binney and Tremaine 1987) is assumed.
Recently,  some new formulae for distribution functions
for both spherical and axisymmetric galaxies have been presented
by Jiang and Ossipkov (2007a,b,c);
but they do not seem applicable to the oblate Jaffe models.

In order to understand the Gauss-Hermite fit parameters clearly, we
first assume that a galaxy is in the frame of reference $(x,y,z)$
with mass density given by Eq.~(\ref{(1.14)}), and
$Ox,y,z$ are its principal axes.
Obviously, $R=\sqrt{x^2+y^2}.$
Then, without loss of
generality,  we take the viewing direction to be in the $(y,z)$ plane
  and
project the galaxy along the line of sight direction, which is taken as
the $z^\prime$-axis, with an  inclination angle $\varphi$ to
the $z$-axis, onto the perpendicular $(x^\prime, y^\prime)$ plane.
Thus a new frame
of reference $(x^\prime, y^\prime, z^\prime)$ is formed by
 the sky plane $(x^\prime, y^\prime)$ and the line of sight which is
 the $z^\prime$-axis;
 the frames of reference
 satisfy the following relations
\begin{equation}
x=x^\prime, \hbox{  }
y=y^\prime \cos \varphi -z^\prime \sin \varphi , \hbox{  }
z=y^\prime\sin \varphi +z^\prime\cos \varphi .
\label{trans}
\end{equation}
When the galaxy is observed at an inclination angle $\varphi,$ the projected
surface density $\Sigma(x^\prime, y^\prime)$
at any point $(x^\prime, y^\prime)$ on the sky plane is
\begin{equation}
\Sigma(x^\prime, y^\prime)=
\int_{-\infty}^{+\infty}\rho([x^\prime]^2+[y^\prime \cos \varphi
-z^\prime \sin \varphi]^2,
 y^\prime\sin \varphi +z^\prime\cos \varphi )dz^\prime.
\label{psd}
\end{equation}

Let $v_{||}$ and $\sigma_{||}$ be the line-of-sight
velocity and velocity dispersion respectively.
The components of velocity directed
along $(x^\prime, y^\prime)$ and
the line of sight $ z^\prime$ can be obtained as follows:
\begin{equation}
v_{x^\prime}=(xv_R-yv_\phi)/R,\hbox{  }
v_{y^\prime}=(yv_R+xv_\phi){\cos \varphi}/{R}+v_z\sin \varphi,
\label{(4.2.4)}
\end{equation}
\begin{equation}
v_{||}\equiv v_{z^\prime}=-(yv_R+xv_\phi)
{\sin \varphi}/{R}+v_z\cos \varphi. \label{(4.2.5)}
\end{equation}

The line-of-sight velocity distribution curve of any
distribution function can be given by evaluating a triple integral of
the distribution function $f(\varepsilon,L_z)$
with respect to  two velocity variables (say, $v_{x^\prime}$ and
$v_{y^\prime}$) and one position variable (say, $z^\prime$)
when $\varepsilon$ and $L_z$ are expressed as functions of six
variables (say, $v_{x^\prime},$ $v_{y^\prime},$ $v_{z^\prime},$ $x^\prime,$
$y^\prime$ and $z^\prime$) in the phase space.
Without loss of
generality, we consider the normalized line-of-sight velocity distribution curve
 \begin{equation}
l_0(v_{||},x^\prime, y^\prime)
=L_0(v_{||},x^\prime, y^\prime)/\Sigma(x^\prime, y^\prime) \label{(4.2.7)}
\end{equation}
with
\begin{equation}
L_0(v_{||},x^\prime, y^\prime)=\int\int\int f(\varepsilon, L_z)d
v_{y^\prime} d v_{x^\prime} d z^\prime, \label{(4.2.6)}
\end{equation}
where the triple integral is to be performed over all the physical values determined by
the potential (see, e.g., Evans 1994).

 On the other hand, any
line-of-sight velocity distribution curve can be quantified by a
Gauss-Hermite series with its Gauss-Hermite coefficients $h_n,$
which can parameterise  the velocity distribution curve to
carry most of information on the shape
of the velocity distribution curve (van der Marel and Franx 1993).
$l_0(v_{||},x^\prime, y^\prime)$ is indeed approximated by
 the following Gauss-Hermite series:
\begin{equation}
l(v_{||},x^\prime, y^\prime)=\hat{\gamma}
\hat{\sigma}^{-1}\sum_{n=0}^\infty h_n u_n(w), \label{(4.2.9)}
\end{equation}
with
\begin{equation}
h_n\equiv h_n(x^\prime, y^\prime)=\frac{
(4\pi)^\frac{1}{2}}{\hat{\gamma}}\int_{-\infty}^{\infty}dv_{||}l_0(v_{||},
x^\prime, y^\prime)u_n(w)
 (n=0,1,2,3,\cdots), \label{(4.2.10)}
\end{equation}
 and
\begin{equation}
u_n(w)=(2^{n+1}\pi n!)^{-\frac{1}{2}}H_n(w)e^{-\frac{w^2}{2}}. \label{(4.2.1)}
\end{equation}
Here $w=(v_{||}-\hat{v})/\hat{\sigma}$ and $H_n(w) (n=0,1,2,\cdots)$
are the standard Hermite polynomials.
Thus the Gauss-Hermite coefficients $h_n$ can be used to
explain  the properties of
the line-of-sight velocity distribution curve of the distribution function of stars
in any elliptical galaxy. $h_0$ represents a Gaussian profile,
and odd and even
Gauss-Hermite coefficients $h_n$ describe asymmetric and
symmetric deviations from a Gaussian profile respectively.
$\hat{v}$ is called the mean radial or streaming velocity,
and $\hat{\sigma}$
the velocity dispersion (van der Marel and Franx 1993).
In fact, the quantities $\hat{v}$ and $\hat{\sigma}$
are two free parameters and differ from
the line-of-sight velocity $v_{||}$ and dispersion $\sigma_{||}.$
$\hat{\gamma}$ is also a free parameter, in addition to
$\hat{v}$ and $\hat{\sigma}.$
The parameter $\hat{\gamma}$ is called the line
strength (van der Marel and Franx 1993).

The Gaussian best fit to the velocity distribution curve is such that
$h_0=1$ and $h_1=h_2=0$ with an  additional condition that the
fit parameter $\hat{\gamma}$ is usually around 1.
The Gaussian best fit parameters $(\hat{v},\hat{\sigma},
\hat{\gamma},h_n)$ can be calculated enabling
comparison with the corresponding observations of elliptical
galaxies.
When the Gaussian best fit parameters $(\hat{v},\hat{\sigma},
\hat{\gamma},h_n)$ are obtained,
improved estimates for the velocity moments can also be given
and so the Gauss-Hermite coefficients $h_n$ are also termed
the Gauss-Hermite moments.
Since $h_0-1=h_1=h_2=0$ and $\hat{\gamma}\simeq 1$ for the above
Gaussian best fit, the other four Gaussian best fit parameters
 $(\hat{v},\hat{\sigma},h_3,h_4)$ determine principally
the velocity distribution curve, and thus
they are usually called the Gauss-Hermite fit parameters.
In fact, once it is assumed that $h_0=1,$ we can regard $\hat{\gamma}$
as a function of the two parameters: $\hat{v}$ and $\hat{\sigma},$
that is, we know from (\ref{(4.2.10)}) that
\begin{equation}
\hat{\gamma}=
\sqrt{2}\int_{-\infty}^{\infty}dv_{||}l_0(v_{||},
x^\prime, y^\prime)e^{-\frac{w^2}{2}}, \label{gamma1}
\end{equation}
which gives that $0<\hat{\gamma}\leq \sqrt{2}$ for each point $(x^\prime,y^\prime).$

\section{Relations between the Model and the Observed Variables}
\label{rmov}
In order to apply the oblate models to analyse a galaxy, we first have to
know the relations between the model variables $(c,d,r_J)$ of Eq.~(\ref{(1.13)})
(or (\ref{(1.14)})) and the physical dimensions of the observed galaxy.
The stellar dynamical model takes
$r_J$  and  $u=\left({GM}/{r_J}\right)^{1/2}$
 as units of length and  velocity, respectively.
Thus, in order to relate $(c,d,r_J)$ in expression (\ref{(1.13)})
 to the physical units of the galaxy, we first have to
change $(c,d,r_J)$ into dimensionless parameters $(\hat{c},\hat{d}),$
assuming that the unit of length is $r_J$ [i.e. using
a transformation $ \hat{c}=cr_J,\hat{d}=dr_J$].
Then a stellar model with dimensionless
parameters $(\hat{c},\hat{d})$ can be obtained [that is, dimensionless
parameters $(\hat{c},\hat{d})$ can be determined from the
relation (\ref{(2.5)})]
in terms of
the image and ellipticity of the galaxy.
Finally, the length and velocity scales $r_J$ and $u$ can be
determined by comparing the Gauss-Hermite parameters of the model
with those of NGC~2974, using a weighted least squares method.
This comparison is made by minimizing a function constructed below in
terms of the weighted least squares method.
Thus we can relate the physical parameters $(c,d,r_J)$
of the models to the physical scales of the galaxy.

Assume that $(\hat{\sigma},\hat{v},h_3,h_4)$ are Gauss-Hermite fit parameters,
 as functions of the dimensionless position variable $x^\prime$
measured from the potential centre along the major axis
for the dimensionless model with variables $(\hat{c},\hat{d}).$
Then $(\hat{\sigma},$ $\hat{v},$ $h_3,$ $h_4)$ are also dimensionless.
Using the linear relation
\begin{equation}
r=r_Jx^\prime, v=u\hat{v},
\sigma=u\hat{\sigma}. \label{(5.2.1)}
\end{equation}
we can relate the physical parameters of the models
with the physical dimensions of the galaxy.

The observations corresponding to the
 Gauss-Hermite fit parameters $(\sigma,$ $v,$ $h_3,$ $h_4)$
are functions of the distance $r$ from the galactic centre.
Generally, $1$ arcsec is chosen as the unit of the distance $r$ and
 $1$ kms$^{-1}$ as the unit of velocity.
The dimensionless model variables $\hat{\sigma}$ and
$\hat{v}$ are functions of dimensionless position $x^\prime.$
For the oblate models the dimensionless coordinate $x^\prime$ is
defined as the distance from the potential centre.
In this paper, it is also
assumed to be a position variable from the potential centre
 along the major axis since we are fitting along the major axis only.
In order to construct a fit
for a galaxy, we have  to
use the linear transformations  (\ref{(5.2.1)}) to convert the dimensionless
model variables $\hat{\sigma},$
$\hat{v}$ and $x^\prime$ into dimensional model variables
 $\sigma,$ $v$ and $r,$ which can be compared with  the
corresponding observations.

From Eq.~(\ref{(5.2.1)}), the Gauss-Hermite coefficients
$h_3$ and $h_4$ in the models can be rewritten as
\begin{equation}
 h_3=h_3(x^\prime)=h_3({r}/{r_J}),
 h_4=h_4(x^\prime)=h_4({r}/{r_J});
\label{mh34}
\end{equation}
and the dimensional model variables
 $\sigma$ and $v$ can be regarded as functions of the distance $r,$ i.e.,
\begin{equation}
v=u\hat{v}\left({r}/{r_J}\right),
\sigma=u\hat{\sigma}
\left({r}/{r_J}\right).
\label{msv}
\end{equation}

Once the length and velocity scales $r_J$ and $u$ have been determined,
 the physical parameters of the models can be
found from Eq.~(\ref{(5.2.1)}),
in order to describe the physical scales of the galaxy.
In order to determine the length and velocity scales
$r_J$ and $u,$
we have to use the weighted least squares method to construct
a function. Then the two scales $r_J$ and $u$
can be obtained by minimizing
the function in the physical domain.
The detailed process of obtaining
the length and velocity scales $r_J$ and $u$ is presented below.

Observations corresponding to the Gauss-Hermite fit parameters
for any galaxy are discrete.
Assume that there are a sequence of observations [say,
$\sigma_i,$ $v_i,$ $h_{3i},$
$h_{4i}$] with errors $(\delta \sigma_i,$ $\delta v_i,$ $
\delta h_{3i},$ $\delta h_{4i}),$
 corresponding to the Gauss-Hermite fit parameters,
at distances $r_i$ $(i=1,2,\cdots,n).$
Let $f(r_J)$ and $g(r_J,u)$ be given by the following two functions:
\begin{equation}
 f(r_J)=\sum_{i=1}^n\left(w_{3i}|h_3({r_i}/{r_J})-h_{3i}|^2
+w_{4i}|h_4({r_i}/{r_J})-h_{4i}|^2\right)
\label{mm01}
\end{equation}
and
\begin{equation}
 g(r_J,u)=\sum_{i=1}^n\left(w_{vi}|u\hat{v}({r_i}/{r_J})-v_i|^2
+w_{\sigma i}|u\hat{\sigma}({r_i}/{r_J})-\sigma_i|^2\right),
\label{mm02}
\end{equation}
where the weights $w_{3i},$ $w_{4i},$ $w_{vi}$ and  $ w_{\sigma i}$ are
related to the observational errors.  Thus
the length and velocity scales $r_J$ and $u$
can be determined by minimizing $f(r_J)+g(r_J,u).$

The scale lengths $r_J$ for the arguments $v$ and $\sigma$ are the same as
those for the arguments $h_n.$ The scale lengths are different
from the effective radius $R_e$ or core radius $R_c.$
Here, the effective radius $R_e$ is defined to be
the radius of the isophote containing
half of the total luminosity and the core radius $R_c$ is where the surface
brightness has fallen to half of its central value.
It is found that the scale length $r_J$ is about half that of
the core radius [i.e. $r_J\simeq R_c/2$] for NGC~2974.

Generally, the weights $w_{3i},$ $w_{4i},$ $w_{vi}$ and  $ w_{\sigma i}$
are chosen as follows.
\begin{equation}
w_{3i}=\frac{e-e^{\delta h_{3i}/h_{3m}}}{e-1}e^{-4\delta h_{3i}/h_{3m}}, ~~
w_{4i}=\frac{e-e^{\delta h_{4i}/h_{4m}}}{e-1}e^{-4\delta h_{4i}/h_{4m}},
\label{ewh34}
\end{equation}
\begin{equation}
w_{vi}=\frac{e-e^{\delta v_i/v_m}}{e-1}e^{-4\delta v_i}, ~~
w_{\sigma i}=\frac{e-e^{\delta \sigma_i/\sigma_m}}{e-1}
e^{-4\delta \sigma_i}.
\label{ewvs}
\end{equation}
We take $h_{3m}=h_{4m}=0.2$ and  $v_m=\sigma_m=300$ kms$^{-1},$
since the range of
the measured velocity dispersions is usually
$100$ -- $300$ kms$^{-1}$ (Gerhard 1993a; Bender et al.~1994)
and the third-order
and fourth-order Gauss-Hermite coefficients
range from $-0.2$ to $0.2$ (Gerhard 1993a; Surma and Bender 1995) for elliptical galaxies.
Our choice of the weights (\ref{ewh34}) and (\ref{ewvs}) is based on
 the following three reasons.
First, the weights should be equal to zero when the
observational errors attain their corresponding
maximal values ($0.2$ for the Gauss-Hermite moments and 300 kms$^{-1}$ for
rotation velocities and velocity dispersions), and unity when
the observational errors are zero. Secondly, the weights should decrease
with the observational errors. Thirdly, the weights
should be continuous functions, for convenience of calculation.

\section{Determination of the Total Mass of the Model}
\label{dtmm}
It has been shown in Sect.~\ref{rmov} that the model
variables can be related to the corresponding observations of
the galaxy if the length and velocity scales
$r_J$ and $u= \left({GM}/{r_J}\right)^{1/2}$ are known.
In this section we show how to use the two scales
$r_J$ and $u$ to calculate the finite total mass $M$ of the
oblate model defined by Eqs.~(\ref{(1.13)}) and (\ref{(1.14)}).

As mentioned in Sect.~\ref{rmov},
  the distance $r$ from the galactic centre is generally
given in  arcsec
(Cinzano and van der Marel 1994; Bender et al.~1994)
and  the model variable $ x^\prime$ is dimensionless.
For convenience when comparing the model variables with the
corresponding observations,
the two scales $r_J$ and $u$ in (\ref{(5.2.1)})
are given in arcsec and kms$^{-1}$ respectively.
Suppose that  $d_0$ is the distance from this galaxy to the observer
in units of Mpc.
Assume that  $p$ (in  arcsec)
is the length scale $r_J $ of a model.
If  the Gauss-Hermite fit parameters for
the  model with the two scales $r_J$ and $u$ can fit
the corresponding observations of a galaxy following
minimization of a function with
(\ref{mm01}) and (\ref{mm02}),
then  $p$ and $u$ can be obtained by comparing
the model variables $(x^\prime, \hat{v}, \hat{\sigma},h_3,h_4)$
with the corresponding observed quantities $(r,v,\sigma,h_3,h_4)$ by use
of the linear transformation (\ref{(5.2.1)}).
On the other hand, the length and velocity units of the model
are $r_J$ and $\left({GM}/{r_J}\right)^{1/2}.$
Since the two scales $r_J$ and $u,$ obtained by comparing the
Gauss-Hermite fit parameters of the models
with the corresponding observations,
are the same as
the length and velocity units of the model respectively,
we have
\begin{equation}
r_J=d_0 \left(\frac{\pi p}{180\times 3600}\right)
\label{length}
\end{equation}
 and
\begin{equation}
\left({GM}/{r_J}\right)^{1/2}=u.
\label{velocity}
\end{equation}
Indeed, (\ref{length}) is also a transformation from a distance unit of $1$
 arcsec to a unit of $1$ Mpc.
Since it has been shown by Jiang (2000) that
the total mass of the model is finite and equal to $M,$
it is known from Eqs.~(\ref{length}) and (\ref{velocity})
that the total mass $M$ of the model is
\begin{equation}
M=\frac{u^2 r_J}{G}=\frac{u^2d_0}{G} \left(\frac{\pi p}
{180\times 3600}\right).
\label{(5.3.1)}
\end{equation}

\section{Fitting of Oblate Jaffe Models}
\label{fojm}
In this section we obtain a good fit of an oblate Jaffe model
by mimicking the image
and ellipticity of the galaxy NGC~2974,
and comparing the Gauss-Hermite fit parameters of
the model with the corresponding observations
of the galaxy. But, it is finally found that the derived
finite total mass of the model is
quite close to the total luminous mass of this galaxy.
\begin{figure}[thb]
\centering
\begin{minipage}[b]{0.48\linewidth}
\centering
\includegraphics[bb=54 641 189 761, width=0.9\linewidth,totalheight=0.8\linewidth]{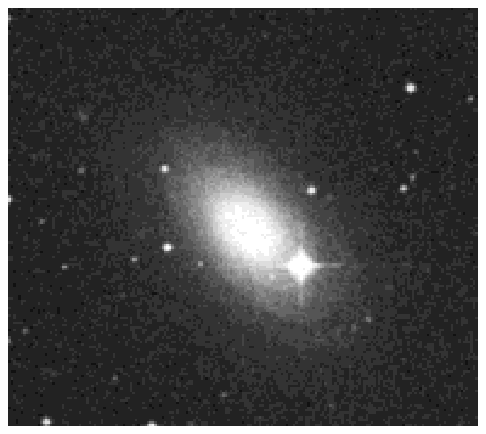}
\caption{The image of NGC~2974 obtained from NASA/IPAC Extragalactic
Database (NED).
\protect\rule[-0.001\baselineskip]{0pt}{0.001\baselineskip} }
\label{ngc2974}
\end{minipage}\vspace*{2mm}
\begin{minipage}[b]{0.48\linewidth}
\centering
\psfrag{z}{$z$}
\psfrag{x}{$R$}
\includegraphics[bb=157 301 436 477, width=\linewidth,totalheight=0.8\linewidth]{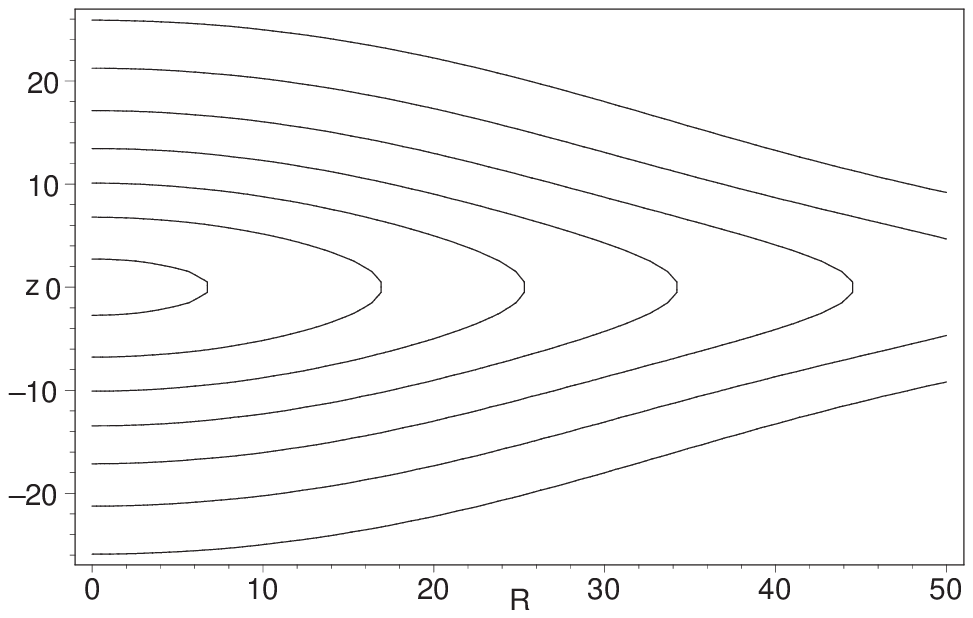}
\caption{The contours of the mass density $\rho(R^2,z)$ in (\ref{(1.14)}) for $c=8.43r_J$ and $d=15r_J.$
Successive contour levels differ by factors of $0.5.$}
%\protect\rule[-0.001\baselineskip]{0pt}{0.001\baselineskip} }
\label{density843c}
\end{minipage}\hspace*{0.02\linewidth}%
\begin{minipage}[b]{0.48\linewidth}
\centering
\psfrag{z}{$z$}
\psfrag{x}{$x$}
\includegraphics[bb=158 299 435 477, width=\linewidth,totalheight=0.8\linewidth]{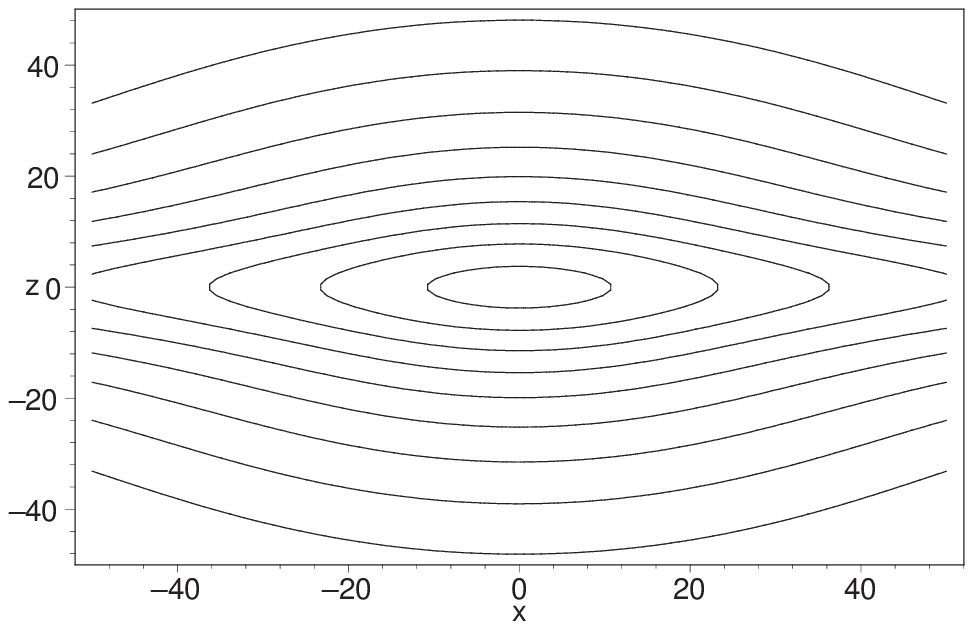}
\caption{The contours of the edge-on projected surface density generated by
$\rho(R^2,z)$ in (\ref{(1.14)}) for $c=8.43r_J$ and $d=15r_J.$
Successive contour levels differ by factors of $0.5.$}
%\protect\rule[-0.001\baselineskip]{0pt}{0.001\baselineskip} }
\label{ds843c}
\end{minipage}
\end{figure}
\begin{figure}[thbp]
\centering
\psfrag{NGC2974 MJ-60}{}
\psfrag{v}{$v$[km/s]}
\psfrag{sigma}{$\sigma$[km/s]}
\psfrag{h3}{$h_3$}
\psfrag{h4}{$h_4$}
\psfrag{x}{$r$[arcsec]}
\includegraphics[bb=58 50 546 740, width=0.8\linewidth]{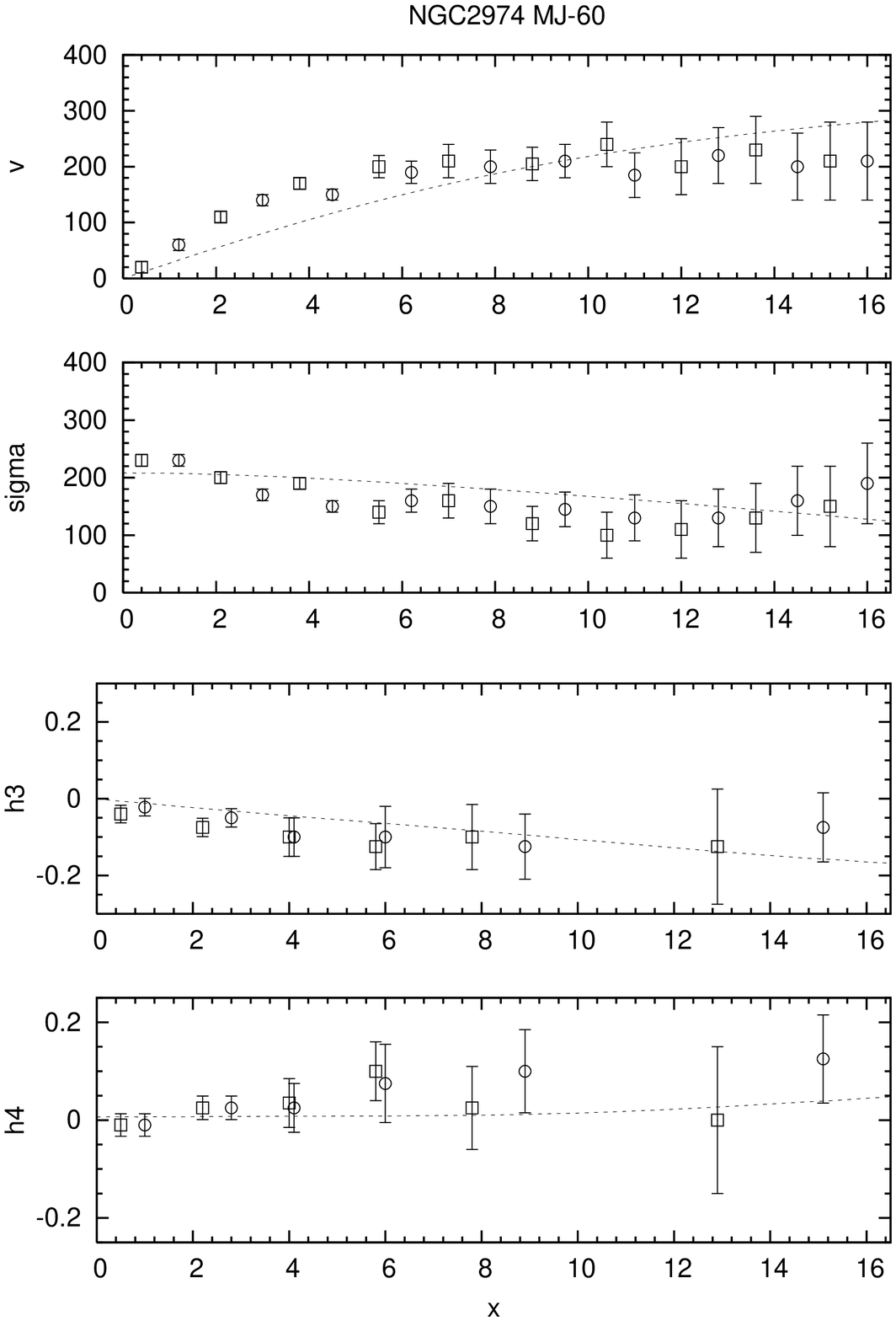}
\caption{Gauss-Hermite fit parameters for NGC~2974: the error bars represent
the observations of Cinzano and van der Marel (1994),
the broken curves come from the
oblate  model defined by Eqs.~(\ref{(1.13)}) and (\ref{(1.14)})
with $c=8.43 r_J$ and $d=15r_J,$ assuming
isotropy with the choice of $r_J=0.55$ arcsec
and $u=3300$  kms$^{-1}.$ The inclination angle of the minor axis to
the line of sight is $60^{\circ}.$}
\label{NGC2974_MJ_60}
\end{figure}

It has been mentioned in Sect.~\ref{rmov} that
the oblate model defined by Eqs.~(\ref{(1.13)}) and (\ref{(1.14)})
with physical parameters $(c,d,r_J)$ first has to be
nondimensionalized  to the dimensionless variables $\hat{c}$
and $\hat{d},$ which  are given by analysing the ellipticity and contours of
the image of the galaxy, so that
the model can mimic the image of the galaxy.
NGC~2974 is classified as E4, and
the flattening of the contours in B--R images of NGC~2974 (Kim 1989)
suggests
that the  inclination angle of the minor axis to the line of sight of
NGC~2974 is about $60^{\circ}$ (Cinzano and van der Marel 1994).
Thus we examine oblate models of ellipticity E4 with
an inclination angle of $60^{\circ}$
to describe NGC~2974. On the other hand, it is easy to see from
Eq.~(\ref{(2.5)}) that
 the  dimensionless model is of type E4 when
$\hat{c}=8.43$ and $\hat{d}=15$ [that is, $c=8.43 r_J$ and $d=15r_J$].
Of course, there are an
infinity of  solutions ($\hat{c},\hat{d}$) of Eq.~(\ref{(2.5)}) that
 make  the  dimensionless model be of type E4, and
$(\hat{c},\hat{d})=(8.43,15)$ is just one of these, and is such that not only is
the ellipticity of the model close enough to that of the galaxy NGC~2974
but also the contours of the projected surface density of the model are quite similar to those
of the image of the galaxy. That is, it can be applied to simulate the image of the galaxy.
We find the image of the galaxy from the NASA/IPAC Extragalactic Database
and plot the contours of the mass density of this model
and its edge-on projected surface density. Here,
Fig.~\ref{ngc2974} shows the image of NGC~2974, and
 Fig.~\ref{density843c} and Fig.~\ref{ds843c}
display the contours of the mass density and its edge-on projected surface
density, respectively. In Fig.~\ref{density843c} and Fig.~\ref{ds843c},
successive contour levels differ by factors of 0.5,
from the potential centre to the outskirts,
and the isophotes get more flattened outwards along the major axis.
This illustates that the isophotes of the model become more disky outwards,
as the mass density drops off slowly like $r^{-3}$ along the major axis.

Once the dimensionless model  is determined by using
the image and ellipticity of the galaxy,
the Gauss-Hermite fit parameters at any inclination angle  for the
model can be calculated, by using the maximum entropy principle
or assuming the isotropy (or anisotropy)
of the model (Jiang 2000).
Then we use a weighted least squares algorithm, as mentioned
in Sect.~\ref{rmov}, to compare the Gauss-Hermite fit parameters
of the model with the corresponding observations of NGC~2974.
Since the model is oblate, we can
assume arbitrarily that the model is isotropic.
Minimization of the function $f(r_J)+g(r_J,u)$ shows that
an oblate model given by Eqs.~(\ref{(1.13)}) and (\ref{(1.14)})
with $c=8.43 r_J$ and $d=15r_J$  can be used to
  determin the Gauss-Hermite fit parameters of NGC~2974,
with an inclination angle of $60^{\circ}.$ Further,
 the length and velocity scales are
$r_J=0.55$ arcsec (or say, $p=0.55$ arcsec) and $u=3300$  kms$^{-1}$
respectively.
   Fig.~\ref{NGC2974_MJ_60} shows this fit.
Our diagnostic or criterion for deciding the quality of the fit
is that the weighted least squares fit is regarded
as good if the mean absolute error values
of the velocity $v,$ the dispersion $\sigma,$
and the Gauss-Hermite coefficients $h_3$ and $h_4$, between the model and the observations, are
less than $15$ kms$^{-1},$  $15$ kms$^{-1},$ $0.01$ and $0.01,$ respectively.
For this model,  the mean absolute error values
between these fit parameters $(v, \sigma, h_3, h_4)$ and their corresponding observations are, respectively,
$12.75$ kms$^{-1},$ $12.00$ kms$^{-1},$ $0.005$ and $0.004.$
Therefore this fitting can be regared as good.

The error bars in Fig.~\ref{NGC2974_MJ_60} represent the major axis
Gauss-Hermite fit parameters of NGC~2974 [see Cinzano and van der
Marel 1994, figure 1 or figure 5(a)]. These observations are
obtained by analysing the galaxy spectrum data with the modified
Fourier fitting method developed  independently by van der Marel and
Franx (1993) and Gerhard (1993b). This data analysis is used to
determine the derivations of the velocity distribution curve from a
Gaussian.

The galaxy spectrum is assumed to be the convolution of a template spectrum and the velocity distribution curve.
The velocity distribution curve is expanded as the Gauss-Hermite series of the five Gauss-Hermite fit parameters,
which are defined by (\ref{(4.2.9)}) with (\ref{(4.2.10)}) and (\ref{(4.2.1)}).
The parameters are determined by $\chi^2$ fitting in
Fourier space of the broadened template spectrum to the galaxy spectrum.
The finite spectral resolution of the instrument needs a high- and a low- wavenumber cut-off.
To obtain parameters ($\gamma,v,\sigma$) from the galaxy spectra, a Gaussian velocity profile can be fitted to
the spectra. This requires less rebinning of the data in the spatial direction.
To get the parameters ($h_3,h_4$), the parametrization (\ref{(4.2.1)})
can be fitted to the data.

The galaxy spectra are in fact the data that are produced by the
standard reduction of stellar spectra with one spectrum of high
signal-to-noise ratio for each star. This data reduction is done as
follows. First, wavelength calibration is done using He-Ar arc lamp
spectra, and so is sky subtraction using the available data at the
ends of the slit. After that, frames should be flat-fielded and
cleaned from cosmic rays and CCD defects. Finally, the galaxy
spectra are rebinned in the spatial direction, with emission lines
of 4959 {\AA}, 5007 {\AA} and 5200 {\AA} interpolated over, in order
to increase the signal-to-noise ratio.

The stellar spectra are obtained by using a spectrograph equipped with CCD.
To do this, one has to consider the exposures of the spectrograph, the scale of the
rebinned stellar spectra, the sky seeing, the dispersion of the grating, the
range of the spectra, the width of the slits, and so on.
The spectroscopic data of slits along the major axis (PA=225$^\circ$) of NGC~2974 are first taken
through two separate 45-min exposures that are reduced separately and added later.
The stellar spectra are then rebinned at the telescope over 3 pixels in the spatial direction and so
have a scale of 1.65 arcsec/pixel.  The seeing during the observations is about 1.9 arcsec
FWHM. The grating used has a dispersion of 0.89 {\AA}/pixel. The chosen spectral range is
from 4750 to 5600 {\AA}, centred at about 5175 {\AA}. The slit width is 1.51 arcsec  so that the
instrument velocity resolution is 45 km/s.

The broken curves in Fig.~\ref{NGC2974_MJ_60} are those for
 the parameters of the oblate model mentioned above.
The curve of $v$ fits well the velocity of NGC~2974
at radii  between 7 to 14 arcsec, although inside a radius of about 7 arcsec,
the model velocity is lower than that observed,
and  is higher at distances larger than about 14 arcsec.
The curve of $\sigma$ lies within the corresponding
error bars in the region from 11 to 16 arcsec,
but it is lower inside a radius of about 2 arcsec
and higher in the range from 2 to 11 arcsec.
As the radius increases,
the dispersion of the model always decreases;
but the dispersion of NGC~2974 tends to increase
with radius from about 12 arcsec.
The graphs of the parameters $h_3$ and $h_4$ are
the best fit for the four broken curves in Fig.~\ref{NGC2974_MJ_60};
they are within, or outside but very close to their corresponding error bars.
In a word, the fit of the four parameters
in Fig.~\ref{NGC2974_MJ_60} is credible, and so it is a good example
to demonstrate our approach of modelling elliptical galaxies dynamically.

The Virgocentric
flow model of Aaronson et al.~(1982) gives
 $d_0=22.5$ h$^{-1}$ Mpc. Eq.~(\ref{length}), when
$p=0.55$ arcsec and $d_0=22.5$ Mpc,
gives $r_J=0.06$ kpc.
Then, by using (\ref{(5.3.1)}),
the total  mass $M$ of the model, given
by $c=8.43 r_J$ and $d=15r_J$, can be calculated as
$M=1.52\times 10^{11}$ $h^{-1}$M$_\odot.$
This  mass is very close to the total luminous mass of NGC~2974
mentioned in Sect.~\ref{intro} [i.e., that
predicted by Cinzano and van der Marel (1994)].
This is consistent with the total
mass $M$
of the models only reflecting the total luminous mass.
This is partly because we choose the model only in terms of
the image and ellipticity of the luminosity of the galaxy,
and partly because the influence of the dark halo in NGC~2974
on the Gauss-Hermite fit parameters of the galaxy
is negligible within a radius of 0.5 $R_{\rm e},$ and the fitting of
Gauss-Hermite fit parameters given in Fig.~\ref{NGC2974_MJ_60}
is made just within this region where
the dynamical influence of the dark halo is negligible
(cf.~also Cinzano and van der Marel 1994).
But the radius of the isophote containing
half of the total mass of this model is $18.59$ arcsec, which
is less than the effective radius of NGC~2974.
This also implies that the dark matter halo extends to
the central area of the ellipsoidal galaxy NGC~2974.
It is worth mentioning that Weijimans et {al.} (2008) recently
found that within 5 $R_{\rm e}$  at least 55 per cent of  the total mass
is dark for the early-type galaxy NGC~2974.
\begin{figure}[thbp]
\centering
\begin{minipage}[b]{0.48\linewidth}
\centering
\psfrag{z}{$z$}
\psfrag{x}{$R$}
\includegraphics[bb=163 297 435 477, width=\linewidth,totalheight=0.8\linewidth]{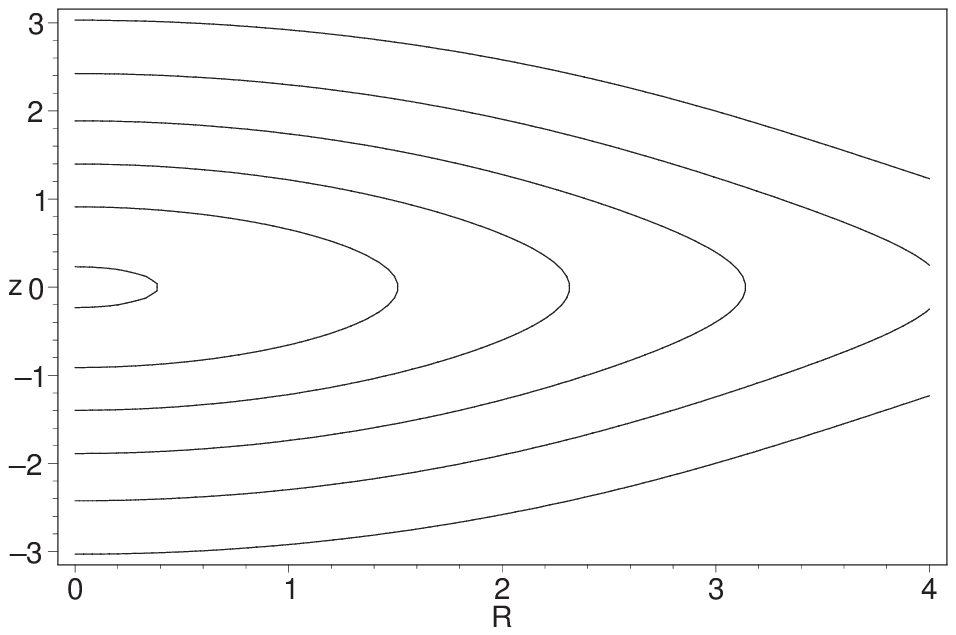}
\caption{As Fig.~\ref{density843c} but for $c=1.26r_J$ and $d=r_J.$}
%\protect\rule[-0.001\baselineskip]{0pt}{0.001\baselineskip} }
\label{density126c}
\end{minipage}\hspace*{0.02\linewidth}%
\begin{minipage}[b]{0.48\linewidth}
\centering
\psfrag{z}{$z$}
\psfrag{x}{$x$}
\includegraphics[bb=163 299 435 477, width=\linewidth,totalheight=0.8\linewidth]{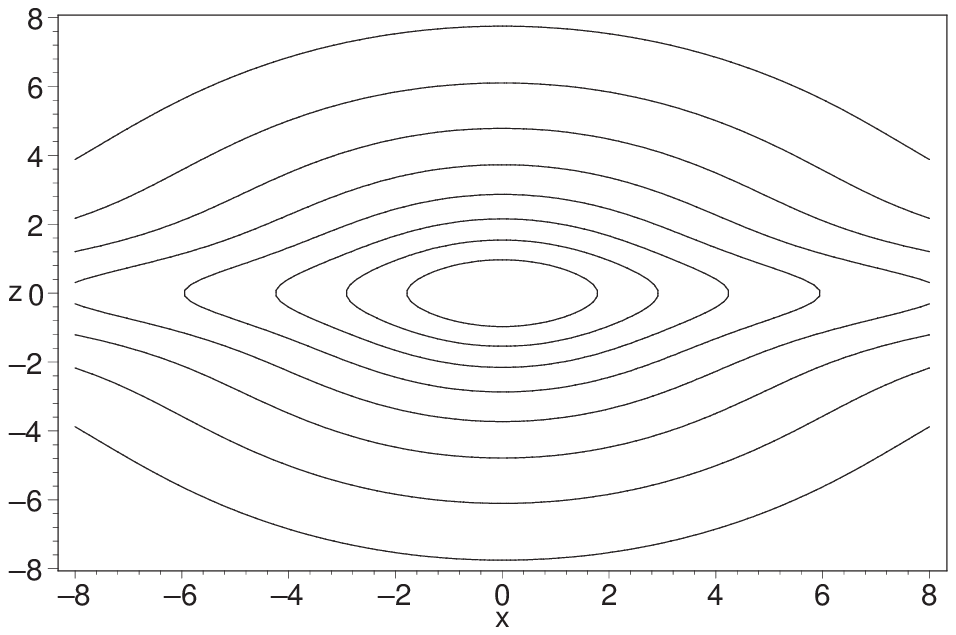}
\caption{As Fig.~\ref{ds843c} but for $c=1.26r_J$ and $d=r_J.$}
%\protect\rule[-0.001\baselineskip]{0pt}{0.001\baselineskip} }
\label{ds126c}
\end{minipage}
\end{figure}
\begin{figure}[thbp]
\centering
\psfrag{NGC2974 MJ-60}{}
\psfrag{v}{$v$[km/s]}
\psfrag{sigma}{$\sigma$[km/s]}
\psfrag{h3}{$h_3$}
\psfrag{h4}{$h_4$}
\psfrag{x}{$r$[arcsec]}
\includegraphics[bb=58 52 532 753, width=0.8\linewidth]{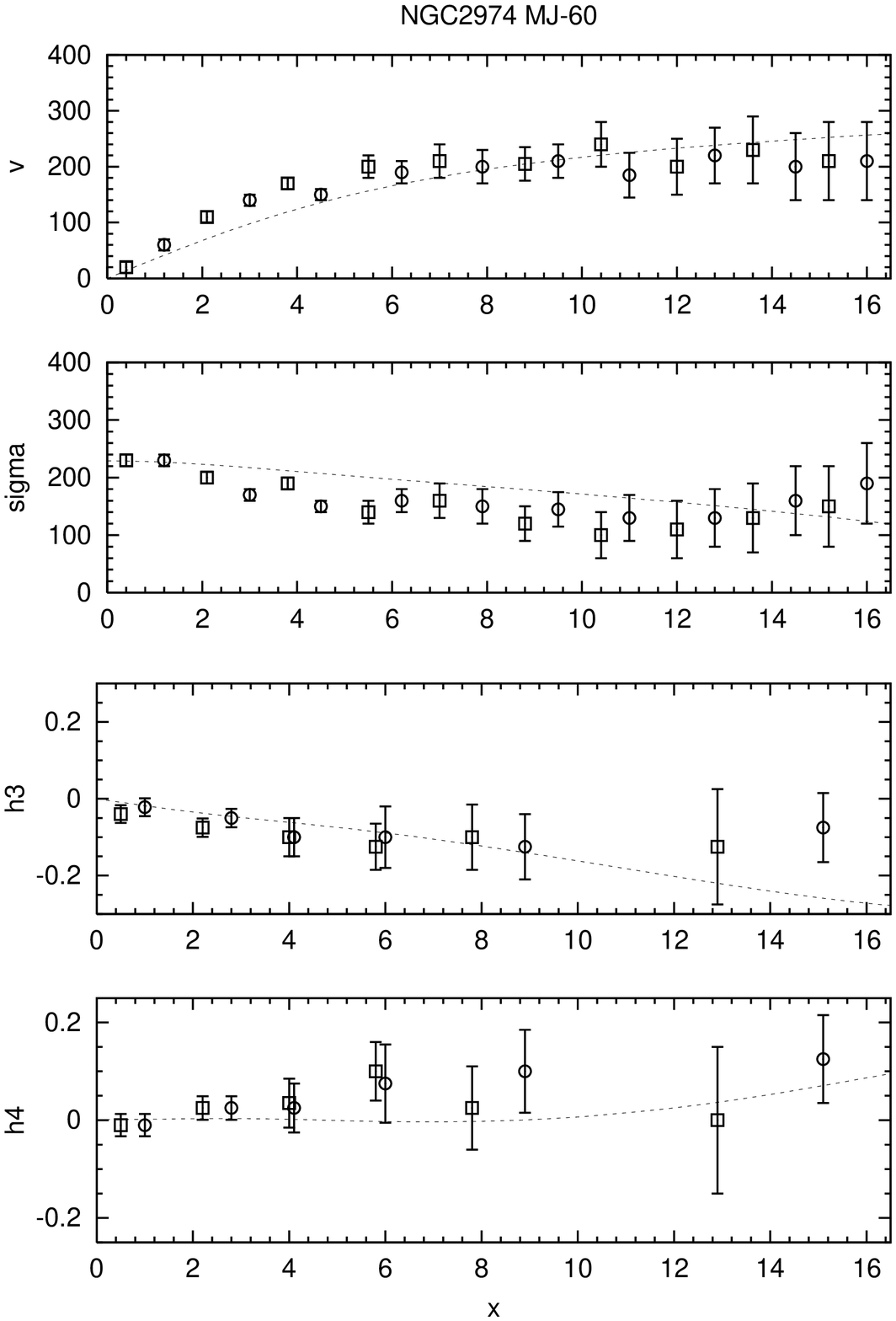}
\caption{As Fig.~\ref{NGC2974_MJ_60}, but the broken curves are obtained for $c=1.26 r_J$ and $d=r_J,$
assuming isotropy, with the length and velocity scales $r_J=3.2$ arcsec and $u=1110$  kms$^{-1}.$
\protect\rule[-0.001\baselineskip]{0pt}{0.001\baselineskip} }
\label{ds126MJ_60}
\end{figure}

Now we consider another model of type E4.  When $(\hat{c},\hat{d})=(1.26,1)$
[that is, $c=1.26 r_J$ and $d=r_J$],  it can be found from Eq.~(\ref{(2.5)}) that
 the  dimensionless model is of type E4.
Fig.~\ref{density126c} and Fig.~\ref{ds126c} show the contours
of the density of this model, and its edge-on projected surface density, respectively.
They are similar to the corresponding figures for $c=8.43 r_J$ and $d=15r_J$
(see Fig.~\ref{density843c} and Fig.~\ref{ds843c}), respectively.
Similarly, in order to
fit the galaxy Gauss-Hermite parameters, the length and velocity scales $(r_J, u)$
can be calculated as follows: $r_J=3.2$ arcsec
(or say, $p=3.2$ arcsec) and $u=1110$  kms$^{-1}.$
Fig.~\ref{ds126MJ_60} shows a Gauss-Hermite parameter fit for NGC 2974.
It can thus be found that the total mass $M$ of this model
is equal to $1.0\times 10^{11}$ $h^{-1}$M$_\odot.$
This mass is less than that of the model with $(\hat{c},\hat{d})=(8.43,15),$ and
it is farther away from the total luminous mass of NGC~2974 than that of the latter.
On the other hand, in Fig.~\ref{NGC2974_MJ_60} and Fig.~\ref{ds126MJ_60},
the corresponding broken curves of the two models
are almost similar and their difference is very small.
Our criterion of determining a better model is thus as follows:
the closer the total mass of the model is to the mass of  NGC~2974, the better the model is.
Therefore, we believe that the model with $(\hat{c},\hat{d})=(8.43,15)$
is a better fit for NGC~2974 than that with $(\hat{c},\hat{d})=(1.26,1),$
although the velocity $v$ fitting in Fig.~\ref{NGC2974_MJ_60} is worse
than that in Fig.~\ref{ds126MJ_60}. For the model with $(\hat{c},\hat{d})=(1.26,1),$
 the mean absolute error values
between the parameters $(v, \sigma, h_3, h_4)$ and their corresponding observations
can be calculated as
$8.75$ kms$^{-1},$ $14.00$ kms$^{-1},$ $0.0085$ and $0.004,$ respectively.
Of course, the fitting in Fig.~\ref{ds126MJ_60} is also a good fit.

In Fig.~\ref{NGC2974_MJ_60} and Fig.~\ref{ds126MJ_60},
the systemic velocity of the two models is monotonic
while the corresponding galaxy observation values flatten out.
This is a shortcoming of our models. It might result from our assumption of isotropy,
which would in turn imply that the galaxy is very anisotropic.

It is further found from our computations that models of type E4
have similar profiles of the
Gauss-Hermite fit parameters $(v,\sigma,h_3,h_4).$
Generally, the smaller the ratio $c/d,$  the better are the models of the type
corresponding to that of the galaxy.
Hence, there might be other
better fit models for NGC~2974, although we cannot discuss these
possibilities here.

\section{Conclusions}
\label{concl}
Dynamical models are generated from both the gravitational potentials and
the distributional functions of the stellar systems, based on dynamical principles.
Therefore the analysis of the dynamical model described above
can be summarised into two steps.
The first is to search for a dimensionless flattened Jaffe model whose
projected surface density generated from its gravitational potential
can be used to fit the actual surface brightness of
the galaxy.   To do this, the ellipticity of the chosen model must be
close enough to that of the image of the galaxy that the contours of
the model are similar to those of the image of the surface brightness of the galaxy.
Of course, the contours of the projected surface density of the model
must become more disky outwards as its corresponding density drops off slowly
along the major axis.
The second step is to determine the length and velocity scales of the
dynamical model by comparing the model variables
with the corresponding observational scales (i.e.~by comparing their
Gauss-Hermite fit parameters
which are needed for quantifying their velocity distribution curves).
The Gauss-Hermite fit parameters of the model are obtained from its distributional functions.

The ellipticity of the model can be determined by the ratio
of $R$-axis to $z$-axis extent of the
contours of the density $\rho(R^2,z)$ near the origin. In order to
compare the fit parameters
$(\hat{v},\hat{\sigma},h_3,h_4)$ of the model with the corresponding
observations, the ellipticity of the model should be the same as that of
the B-R images of the considered galaxy.
The ellipticity of the galaxy being considered
 can be obtained from observations.
By using the  ellipticity obtained, a model can be chosen to fit the image of the
surface brightness of the galaxy
being considered. For example, for the oblate model,
 $\hat{c}$ can be obtained from Eq.~(\ref{(2.5)})
if it is assumed that $c=\hat{c}r_J$ and
that $d=r_J.$
 The model has first to be transformed into a dimensionless
model in terms of $\hat{c},$ and then the fit parameters
$(\hat{v},\hat{\sigma},h_3,h_4)$ of the dimensionless model
have to be calculated. It is necessary to make an assumption
about the anisotropy of the models; we choose an isotropic model.

Comparison with observations requires relations between
the observational scales
and the model variables. The relations are given by the linear
transformation
(\ref{(5.2.1)}). This transformation can be determined if $r_J$
and $({GM}/{r_J})^{1/2}$
are known. Therefore, comparing the model variables with observations by
use of the linear transformation, a
good pair of values of $r_J$ and $({GM}/{r_J})^{1/2}$
can be chosen, such that
the model variables fit the corresponding observations.

The total mass of the flattened Jaffe models can be shown to be $M.$
Thus the values of $r_J$ and $({GM}/{r_J})^{1/2}$
finally lead to the total mass
$M$ of the model.  Since the model only reflects the B-R images of the
galaxy being considered, namely NGC~2974,
the total mass $M$ of the model is obviously less than the total mass of
this galaxy and so it reflects the total luminous mass of this galaxy.
Additionally,
 the dynamical influence of dark matter in NGC~2974 on
 the Gauss-Hermite fit parameters
of the galaxy observed is negligible inside a radius of $0.5$ $R_{\rm e}.$

For a very flattened model, the two values of $r_J$ and $({GM}/{r_J})^{1/2}$
can be also determined by comparing the circular velocities generated
from the potential of the model with data of the rotation velocities of the
very flattened galaxies observed. This is the customary approach of
fitting the projected surface density of the model to the observed surface brightness
of the galaxies. This result must be the same as that obtained
by comparing the Gauss-Hermite fit parameters, if the model is a very good fit.

The resulting fit in Fig.~\ref{NGC2974_MJ_60} is satisfactory, and
so it is a good example to illustrate our method for constructing such models.
And, up to now, no other plausible alternatives have been found
for choosing a suitable flattened Jaffe model
for the elliptical galaxy NGC~2974.

\begin{acknowledgement}
ZJ is supported by NSFC 10271121 and by joint grants
of NSFC 10511120278/10611120371 and RFBR 04-02-39026.
The considerable
assistance given by  members of
the High Performance Computing Service of Manchester Computing
is acknowledged.
ZJ is indebted to Dr Paul Stewart, Dr Richard James,
Dr N.~Wyn Evans and Dr R.~P.~van der Marel
for helpful comments.
ZJ thanks Professor Joan Walsh for her
advice on data fitting and Professor Christopher Hunter for
his advice on Hunter and Qian's algorithm (1993).
ZJ is also very grateful to Professor Konstantin Kholshevnikov,
Sergei Kutuzov, Vadim Antonov and Dr Leonid Ossipkov
for their valuable discussions on this work.
ZJ thanks Dr David Moss very much for his providing a lot of scientific ideas in this work.
Finally, ZJ would like to thank the referee of this paper for his/her
helpful comments and suggestions.
\end{acknowledgement}


%\vspace*{0.17cm}
%\newpage
{\noindent\bf References}\vspace*{0.1cm}
{\small
\begin{description}
\item[Aaronson M., Huchra J.~P., Mould J., Schechter P.~L., Tully R.~B.:]  The velocity field in the local supercluster. Astrophysical Journal, {\bf 258}, 64-76 (1982).
\item[Baes M., Dejonghe H.:] A completely analytical family of dynamical models for spherical galaxies and bulges with a central black hole.
Monthly Notices of the Royal Astronomical Society, {\bf 351}, 18-30 (2004).
\item[Baes M., Dejonghe H.,  Buyle P.:] The dynamical structure of isotropic spherical galaxies with a central black hole.
   Astronomy \& Astrophysics, {\bf 432}, 411-422 (2005)..
\item[Bender R., Saglia R.~P., Gerhard O.~E.:] Line-of-Sight Velocity Distributions of Elliptical Galaxies.
Monthly Notices of the Royal Astronomical Society, {\bf 269},
785-813 (1994).
\item[Bennett C.~L., et al.:] First-Year Wilkinson Microwave Anisotropy Probe
(WMAP) Observations: Preliminary Maps and Basic Results. The
Astrophysical Journal Supplement Series, {\bf 148}, 1-27 (2003).
\item[Bennett C.~L., et al.:] The Microwave Anisotropy Probe
Mission. The Astrophysical Journal, {\bf 583}, 1-23 (2003).
\item[Binney J., Tremaine S.:] Galactic Dynamics. Princeton Univ., Princeton (1987).
\item[Canizares C.~R., Fabbiano G., Trinchieri G.:] Properties of the X-ray emitting gas in early-type galaxies. Astrophysical Journal, {\bf 312}, 503-513 (1987).
\item[Cappellari M., Emsellem E.:] Parametric Recovery of Line-of-Sight Velocity Distributions from Absorption-Line Spectra of Galaxies via Penalized
Likelihood. The Publications of the Astronomical Society of the
Pacific, {\bf 116}, 138-147 (2004).
\item[Cinzano P., van der Marel R.~P.:] Observations and Dynamical Modelling of the e4 Galaxy NGC2974 - Evidence for an Embedded Stellar Disc.
Monthly Notices of the Royal Astronomical Society, {\bf 270}, 325-340 (1994).
\item[Davies R.~L., Burstein D., Dressler A., Faber S.~M., Lynden-Bell D., Terlevich R., Wegner G.:]
Spectroscopy and photometry of elliptical galaxies. II - The spectroscopic parameters. Astrophysical Journal, {\bf 64}, 581-600 (1987).
\item[Dehnen W.:] A family of potential-density pairs for spherical galaxies and bulges.
Monthly Notices of the Royal Astronomical Society, {\bf 265}, 250-256 (1993).
\item[Dejonghe H.:] Stellar dynamics and the description of stellar systems.
Physics Reports, {\bf 133}, 217-313 (1986).
\item[Demoulin-Ulrich M.~H., Butcher H.~R., Boksenberg A.:] Extended gaseous emission in normal elliptical galaxies.
Astrophysical Journal, {\bf 285}, 527-546 (1984).
\item[de Vaucouleurs G., de Vaucouleurs A., Faber S.~M., Jr Buta R.~J., Paturel H.~G., Fouqu\'{e} P.:] Reference Catalogue of Bright Galaxies, Springer-Verlag, New York, 3rd (1991).
\item[Emsellem E., Goudfrooij P., Ferruit P.:] A two-arm gaseous spiral in the inner 200 pc of the early-type galaxy NGC 2974: signature of an inner
bar. Monthly Notices of the Royal Astronomical Society, {\bf 345},
1297-1312 (2003).
\item[Evans N.~W.:] The power-law galaxies. Monthly Notices of the Royal Astronomical Society, {\bf 267}, 333-360 (1994).
\item[Forman W., Jones C., Tucker W.:] Hot coronae around early-type galaxies. Astrophysical Journal, {\bf 293}, 102-119 (1985).
\item[Gallouet L., Heidmann N., Dampierre F.:]  Optical positions of bright galaxies. III.
Astronomy \& Astrophysics Supplement Series, {\bf 19}, 1-19 (1975).
\item[Gerhard  O.~E.:] Elliptical Galaxies, in Lecture notes in physics, Vol.~433 (1993a).
\item[Gerhard  O.~E.:] Line-of-sight velocity profiles in spherical galaxies:
breaking the degeneracy between anisotropy and mass.
Monthly Notices of the Royal Astronomical Society, {\bf 265}, 213-230 (1993b).
\item[Gerhard  O.~E., Jeske G., Saglia R.~P., Bender R.:] Breaking the degeneracy between anisotropy and mass:
the dark halo of the E0 galaxy NGC~6703. Monthly Notices of the
Royal Astronomical Society, {\bf 295}, 197-215 (1998).
\item[Halliday C., Davies R.~L., Kuntschner H., Birkinshaw M., Bender R., Saglia R.~P.,  Baggley G.:]
Line-of-sight velocity distributions of low-luminosity elliptical galaxies.
Monthly Notices of the Royal Astronomical Society, {\bf 326}, 473-489 (2001).
\item[Hunter C., Qian E.:] Two-integral distribution functions for axisymmetric galaxies.
Monthly Notices of the Royal Astronomical Society, {\bf 262}, 401-428 (1993).
\item[Jaffe W.:] A simple model for the distribution of light in spherical galaxies.
Monthly Notices of the Royal Astronomical Society, {\bf 202}, 995-999 (1983).
\item[Jiang Z.:] Flattened Jaffe models for galaxies. Monthly Notices of the Royal Astronomical Society, {\bf 319}, 1067-1078 (2000).
\item[Jiang Z., Fang D., Liu H., Moss D.:] General flattened Jaffe models for galaxies. AMS/IP Studies in Advanced Mathematics, {\bf 29},
Geometry and Nonlinear Partial Differential Equations, {p.} 31-37 (2002).
\item[Jiang Z., Fang D., Moss D.:] Axisymmetric models for galaxies by equipotential and equidensity methods.
Proceedings of the Sixth Conference of China Society
for Industry and Applied Mathematics,
Research Information Ltd, Hertfordshire, UK, {p.} 79-83 (2002).
\item[Jiang Z., Moss D.:] Prolate Jaffe models for galaxies.
Monthly Notices of the Royal Astronomical Society, {\bf 331}, 117-125 (2002).
\item[Jiang Z., Ossipkov L.~P.:] Flattened $\gamma$ models for galaxies. Astronomical and Astrophysical
Transactions, {\bf 25}, 213-216 (2006).
\item[Jiang Z., Ossipkov L.~P.:] Anisotropic distribution functions for spherical galaxies.
Celestial Mechanics and Dynamical Astronomy, {\bf 97}, 249-265 (2007a).
\item[Jiang Z., Ossipkov L.~P.:] Two-integral distribution functions for axisymmetric systems.
Monthly Notices of the Royal Astronomical Society, {\bf 379}, 1133-1142 (2007b).
\item[Jiang Z., Ossipkov L.~P.:] Two-integral distribution functions for axisymmetric stellar systems with separable densities.
 Monthly Notices of the Royal Astronomical Society, {\bf 382}, 1971-1981 (2007c).
\item[Kim D. W.:] Interstellar matter in early-type galaxies - Optical observations. Astrophysical Journal, {\bf 346}, 653-674 (1989).
\item[Kim D. W., Guhathakurta P., van Gorkom J.~H., Jura M., Knapp G.~R.:] H I observations of the elliptical galaxies NGC 2974 and NGC 5018.
Astrophysical Journal, {\bf 330}, 684-694 (1988).
\item[Krajnovi\'{c} D., Cappellari M., Emsellem E., McDermid R.~M., de Zeeuw
P.~T.:] Dynamical modelling of stars and gas in NGC 2974:
determination of mass-to-light ratio, inclination and orbital
structure using the Schwarzschild method. Monthly Notices of the
Royal Astronomical Society, {\bf 357}, 1113-1133 (2005).
\item[Kronawitter A., Gerhard O., Saglia R.~P., Bender R.:]
Dynamical analysis of elliptical galaxy halos.
Galaxy Dynamics, ASP Conference Series, {\bf 182}, 441-442 (1999).
\item[Kutuzov S.~A., Ossipkov L.~P.:] A generalized model for the three-dimensional gravitational potential of stellar systems.
Pis'ma v Astronomicheskij Zhurnal, {\bf 57}, 28-37 (1980) (English translations:
Soviet Astronomy Letters, {\bf 24}, 17-22, (1981)).
\item[Kuzmin G.~G., Veltmann \"{U}.-I.K.:] Hydrodynamic models of spherical stellar systems.
 W.~Struve Tartu Astrof\"u\"us.~Obs.~Publ., {\bf 36}, 3-47 (1967a).
\item[Kuzmin G.~G., Veltmann \"{U}.-I.K.:] Lindblad diagram and isochronic
      models. W.~Struve Tartu Astrof\"u\"us.~Obs.~Publ., {\bf 36}, 470-507 (1967b).
\item[Kuzmin G.~G., Veltmann \"{U}.-I.K.:] Generalized isochrone models for
      spherical stellar systems.
      Dynamics of Galaxies and Star Clusters, Nauka, Alma-Ata, 82-87 (1973)
      (English translations: Galactic Bulges (IAU Symp. 153), Kluwer, Dordrecht,
       363-366 (1993)).
\item[Kuzmin G.~G., Veltmann \"{U}.-I.K., Tenjes P.~L.:]  Quasi-isothermal
      models of spherical stellar systems. Application to the galaxies M~87 and
      M~105.
      W.~Struve Tartu Astrof\"{u}\"{u}s.~Obs. Publ., {\bf 51}, 232-242 (1986).
\item[Ossipkov L.~P.:] Some problems of the theory of self-consistent models for
       star clusters.
       Star Clusters, Urals Univ.~Press, Sverdlovsk, 72-89 (1979a).
\item[Ossipkov L.~P.:] Spherical systems of gravitating bodies with an
      ellipsoidal velocity distribution.
      Pis'ma v Astronomicheskij Zhurnal, {\bf 5}, 77-80 (1979b)
      (English translations: Soviet Astronomy Letters, {\bf 5}, 42-44).
\item[Ossipkov L.~P., Jiang Z.:] Density asymptotics for infinite gravitating system.
Messenger of Saint Petersburg University, {Ser.} {\bf 10}, Applied Mathematics, Informatics, Control Processes,
{Iss.} {\bf 2}, 66-74 (2007a).
\item[Ossipkov L.~P., Jiang Z.:] Constructing galaxy models with a central density cusp by equipotential method.
Messenger of Petersburg University, {Ser.} {\bf 1}, Mathematics, Mechanics, Astronomy, {Iss.} {\bf 1}, 139-143 (2007b).
\item[Qian E. E., de Zeeuw P. T., van der marel R. P., Hunter C.:] Axisymmetric galaxy models with central black holes, with an application to M32.
Monthly Notices of the Royal Astronomical Society, {\bf 274},
602-622 (1995).
\item[Schwarzschild M.:] A numerical model for a triaxial stellar system in dynamical
equilibrium. Astrophysical Journal, {\bf 232}, 236-247 (1979).
\item[Saglia R. P., Kronawitter A., Gerhard O., Bender R.:]
The orbital structure and potential of NGC~1399.
Astronomical Journal, {\bf 119}, 153-161 (2000).
\item[Saha P.:] Designer basis functions for potentials in galactic dynamics.
Monthly Notices of the Royal Astronomical Society, {\bf 262}, 1062-1064 (1993).
\item[Sandage A.~R., Tammann G.~A.:] Revised Shapley-Ames Catalogue of Galaxies, Washington DC: Carnegie Institution of Washington (1981).
\item[Surma P., Bender R.:] Relics of dissipational merging and past violent starbursts in elliptical galaxies - the gE galaxy NGC 4365.
Astronomy \& Astrophysics, {\bf 298}, 405-419 (1995).
\item[Thomas, J.~et al.:]
Dynamical modelling of luminous and dark matter in 17 Coma early-type galaxies.
Monthly Notices of the Royal Astronomical Society, {\bf 382}, 657-684 (2007).
\item[Tremaine S.~et al.:] A family of models for spherical stellar systems.
Astronomical Journal, {\bf 107}, 634-644 (1994).
\item[van der Marel R. P., Franx M.:] A new method for the identification of non-Gaussian line profiles in elliptical galaxies.
Astrophysical Journal, {\bf 407}, 525-539 (1993).
\item[Veltmann \"{U}.-I.K.:] Constructing models for spherical star systems with
      given space density.
      Tartu {Astron.~}{Obs.~}Publ., {\bf 33}, 387-415 (1961).
\item[Veltmann \"{U}.-I.K.:] On phase density of spherical stellar systems.
      {W.~}Struve Tartu {Astrof\"{u}\"{u}s.~}{Obs.~}Publ., {\bf 35}, 5-26 (1965).
\item[Veltmann \"{U}.-I.K.:] Phase space  models for star clusters.
       Star Clusters, Urals {Univ.~}Press, Sverdlovsk,  50-71 (1979).
\item[Veltmann \"{U}.-I.K.:] Gravitational potential, space density and phase
      density of star clusters.
      W.~Struve Tartu Astrof\"{u}\"{u}s.~Obs. Publ., {\bf 48}, 232-261 (1981).
\item[Weijmans A., Krajnovi\'{c} D., van de Ven G., Oosterloo T.~A., Morganti R., de Zeeuw P.~T.:]
The shape of the dark matter halo in the early-type galaxy NGC~2974.
Monthly Notices of the Royal Astronomical Society, {\bf 383}, 1343-1358 (2008).

\end{description}
}
\end{document}